\documentclass[12pt]{book}

\usepackage[dvips]{graphicx,color}
\usepackage{makeidx,gamma}

\makeauthorindex
\makeindex

\BookTitle{Physics with GeV Electrons and Gamma-Rays}
\CopyRight{\copyright 2001 by Universal Academy Press, Inc.}

\begin{document}

\BookTitle{\itshape Physics with GeV Electrons and Gamma-Rays}
\CopyRight{\copyright 2001 by Universal Academy Press, Inc.}
\pagenumbering{arabic}

\chapter{
Mesons electromagnetic production study
{\it via} a constituent quark approach}

\author{%
Bijan Saghai\\
{\it Service de Physique Nucl\' eaire, DAPNIA, CEA/Saclay,
91191 Gif-sur-Yvette, France, bsaghai@cea.fr}}
%
%
\AuthorContents{Bijan Saghai} 
\AuthorIndex{Saghai}{B.} 

\section*{Abstract}
Using a chiral constituent quark approach based on the 
broken $SU(6)\otimes O(3)$ symmetry,
we focus on the spectroscopy of isospin-1/2 nucleonic resonances. 
A model for the $\eta$ photoproduction, embodying all known 
nucleonic resonances,
shows clear need for a yet undiscovered $S_{11}$ resonance, 
for which we determine the mass (1.730 GeV) and 
the total width (180 MeV).
\section{Introduction}
The advent of new facilities 
offering high quality electron and photon beams and 
sophisticated detectors, has stimulated intensive experimental
and theoretical study of the mesons photo- and electro-production.

Among various formalisms~[1], the advantage of the quark model in 
this realm  is 
two fold : i) it allows us to embody, in the reaction mechanism, all known 
baryonic resonances,
ii) the electromagnetic production data can directly be related to the 
internal structure of those resonances. As a result, such an approach
offers a reliable frame in search for new resonances.

Within a constituent quark model based on the $SU(6)\otimes O(3)$ symmetry,
we have investigated~[1-6]
the following reactions:
\begin{eqnarray}
{\gamma}~p~& {\rightarrow} &~\eta~ p\label{g},~K^+ \Lambda, 
K^+ \Sigma^\circ\label{Y},~\Phi~ p\label{Phi} \\
e ~ p~& \rightarrow &~e' ~ {\eta}~ p\label{e}.
\end{eqnarray}

Here, we focus on the $\eta$ photoproduction, for which rather copious recent data 
are available and more is expected in the near future.
\section{Theoretical frame}
The starting point of the meson electromagnetic production in the chiral quark 
model is the low 
energy QCD Lagrangian~[7-8].
The baryon resonances in the {\it s-} and {\it u-}channels are treated as three
quark systems.
The transition matrix elements based on the low energy QCD
Lagrangian include the {\it s-}, {\it u-}, and {\it t-}channel contributions
\begin{equation}\label{eq:Msu}
{\cal M}_{if}={\cal M}_s+{\cal M}_{u}+{\cal M}_{t}.
\end{equation}
The contributions from  the {\it s-}channel resonances can be written as
\begin{eqnarray}\label{eq:MR}
{\mathcal M}_{N^*}=\frac {2M_{N^*}}{s-M_{N^*}(M_{N^*}-i\Gamma(q))}
e^{-\frac {{k}^2+{q}^2}{6\alpha^2_{ho}}}{\mathcal A}_{N^*},
\end{eqnarray}
where  $k$ and $q$ represent the momenta of the incoming photon 
and the outgoing meson respectively, $\sqrt {s}\equiv W$ is the total c.m. energy of 
the system, $e^{- {({k}^2+{q}^2)}/{6\alpha^2_{ho}}}$ is a form factor 
in the harmonic oscillator basis with the parameter $\alpha^2_{ho}$ 
related to the harmonic oscillator strength in the wave-function, 
and $M_{N^*}$ and $\Gamma(q)$ are the mass and the total width of 
the resonance, respectively.  The amplitudes ${\mathcal A}_{N^*}$ 
are divided into two parts~[8]: the contribution 
from each resonance below 2 GeV, the transition amplitudes of which 
have been translated into the standard CGLN amplitudes in the harmonic 
oscillator basis, and the contributions from the resonances above 2 GeV
treated as degenerate, since little experimental information is available
on those resonances.

The {\it u-}channel contributions are divided into the nucleon Born
term and the contributions from the excited resonances.  The matrix 
elements for the nucleon Born term is derived explicitly, while the 
contributions from the excited resonances above 2 GeV for a given parity 
are assumed to be degenerate so that their contributions could be 
written 
in a compact form.

The {\it t-}channel contribution contains two parts: 
{\it i)} charged meson exchanges which are proportional to the charge of outgoing 
mesons and thus do not contribute to the process $\gamma N\to \eta N$;  
{\it ii)} $\rho$- and $\omega$-exchange in the $\eta$ production which are 
excluded here due to the duality hypotheses.
\begin{table}[t]
    \caption{Resonances included explicitly in our study with their 
assignments in $SU(6)\otimes O(3)$ configurations, masses, 
and widths. The mass and width of the $S_{11}(1535)$ resonance,
left as adjustable parameters, are given in Table 2.
Higher mass resonances are treated as degenerate.} 
\vspace{.5pc}
\begin{center}
\begin{tabular}{llccccc}  
States & $SU(6)\otimes O(3)$& Mass (GeV)& Width (MeV)  \\ \hline        
$S_{11}(1535)$&$N(^2P_M)_{\frac 12^-}$& & \\ 
$S_{11}(1650)$&$N(^4P_M)_{\frac 12^-}$&1.650&150 \\    
$D_{13}(1520)$&$N(^2P_M)_{\frac 32^-}$&1.520&130\\    
$D_{13}(1700)$&$N(^4P_M)_{\frac 32^-}$&1.700&150\\
$D_{15}(1675)$&$N(^4P_M)_{\frac 52^-}$&1.675&150\\
$P_{13}(1720)$&$N(^2D_S)_{\frac 32^+}$&1.720&150\\    
$F_{15}(1680)$&$N(^2D_S)_{\frac 52^+}$&1.680&130\\    
$P_{11}(1440)$&$N(^2S^\prime_S)_{\frac 12^+}$&1.440&150\\    
$P_{11}(1710)$&$N(^2S_M)_{\frac 12^+}$&1.710&100\\    
$P_{13}(1900)$&$N(^2D_M)_{\frac 32^+}$&1.900&500\\    
$F_{15}(2000)$&$N(^2D_M)_{\frac 52^+}$&2.000&490\\ \hline
\end{tabular}
\end{center}
\end{table}

Within
the exact $SU(6)\otimes O(3)$ symmetry the $S_{11}(1650)$ and 
$D_{13}(1700)$ do not contribute to the investigated 
reaction mechanism. However, the breaking of this symmetry leads to
the configuration mixings.

Here, the most relevant configuration mixings are those of the
two $S_{11}$ and the two $D_{13}$ states around 1.5 to 1.7 GeV. The 
configuration mixings, generated by the gluon exchange interactions in 
the quark 
model~[9-10], can be expressed in terms of the mixing angle
between the two $SU(6)\otimes O(3)$ states $|N(^2P_M)>$  and 
$|N(^4P_M)>$, with the total quark spin 1/2 and 3/2;  
%
%
\begin{eqnarray}\label{eq:MixS}
\left(\matrix{|S_{11}(1535)> \cr
|S_{11}(1650)>\cr}\right) &=&
\left(\matrix{ \cos \Theta _{S} & -\sin \Theta _{S}\cr
\sin \Theta _{S} & \cos \Theta _{S}\cr}\right) 
 \left(\matrix{|N(^2P_M)_{{\frac 12}^-}> \cr
|N(^4P_M)_{{\frac 12}^-}>\cr}\right),  
\end{eqnarray}
%
%
and  
%
%
\begin{eqnarray}\label{eq:MixD}
\left(\matrix{|D_{13}(1520)> \cr
|D_{13}(1700)>\cr}\right) &=&
\left(\matrix{ \cos \Theta _{D} & -\sin \Theta _{D}\cr
\sin \Theta _{D} & \cos \Theta _{D}\cr}\right)
\left(\matrix{|N(^2P_M)_{{\frac 32}^-}> \cr
|N(^4P_M)_{{\frac 32}^-}>\cr}\right),  
\end{eqnarray}
%
%
 
The amplitudes ${\mathcal A}_{N^*}$ in terms of the 
product of the photo- and meson-transition amplitudes
are related to the mixing angles,
\begin{eqnarray}\label{eq:MixAR}
{\mathcal A}_{N^*} \propto <N|H_m| N^*><N^*|H_e|N>,
\end{eqnarray}
where $H_m$ and $H_e$ are the meson and photon transition operators,
respectively. Using Eqs.~(\ref{eq:MixS}) to~(\ref{eq:MixAR}), 
for the resonance ${S_{11}(1535)}$ one finds
\begin{eqnarray}\label{eq:MixAS1}
{\mathcal A}_{S_{11}} &\propto& 
<N|H_m (\cos \Theta _{S}
 |N(^2P_M)_{{\frac 12}^-}> - 
\sin \Theta _{S} 
|N(^4P_M)_{{\frac 12}^-}>) 
 (\cos \Theta _{S} <N(^2P_M)_{{\frac 12}^-}| -
\nonumber\\
&&
\sin \Theta _{S} <N(^4P_M)_{{\frac 12}^-}|)  
 H_e|N>,
\end{eqnarray}
Due to the Moorhouse selection rule,
the photon transition amplitude $<N(^4P_M)_{{\frac 12}^-}|H_e|N>$
vanishes in our model.
So, Eq.~(\ref{eq:MixAS1}) becomes
\begin{eqnarray}\label{eq:MixAS2}
{\mathcal A}_{S_{11}}&\propto& (\cos^2 \Theta _{S} - {\cal {R}}
\sin 
\Theta _{S}\cos \Theta _{S}) 
<N|H_m|N(^2P_M)_{{\frac 12}^-}>
<N(^2P_M)_{{\frac 12}^-}|H_e|N>,
\end{eqnarray}
where $<N|H_m|N(^2P_M)_{{\frac 12}^-}> <N(^2P_M)_{{\frac 12}^-}|H_e|N>$
determines the CGLN amplitude for the 
$|N(^2P_M)_{{\frac 12}^-}> $ state, and the ratio
${\cal {R}} =   {<N|H_m|N(^4P_M)_{{1/2}^-}>}/
{<N|H_m|N(^2P_M)_{{1/2}^-}>}$
is
a constant determined by the $SU(6)\otimes O(3)$ symmetry. Using the 
meson transition operator $H_m$ from the Lagrangian used in deriving 
the CGLN amplitudes in the quark model, we find ${\cal {R}}$~=~-1 for the $S_{11}$
and $\sqrt{1/10}$ for the $D_{13}$ resonances.
\section{Results}
Using the above approach, we have fitted the following sets of the 
$\eta$-photoproduction data: 
differential cross-sections from
MAMI/Mainz~[11] and Graal~[12], as well as the polarized beam asymmetry from 
Graal~[13]. Then we have predicted~[1] the total cross-section and the polarized
target asymmetry. This latter observable has been measured at ELSA/Bonn~[14].
%
\begin{table}[b]
\caption{Results of minimizations.}
\begin{center}
\begin{tabular}{c|cc|cc}
{ parameter}  & \multicolumn{1}{c}{Model I}  & \multicolumn{1}{c|}{Model II }
 & \multicolumn{1}{c}{Isgur-Karl~[9] } & \multicolumn{1}{c}{} \\
\hline
$\Theta_{S}$ (deg.)& -32~$\pm$~2 & -27~$\pm$~1 & -32 &  \\
$\Theta_{D}$ (deg.) &~~~5.1~$\pm$~0.2 &~~~5.1~$\pm$~0.2 & ~~6 \\[1ex]
$M_{S_{11}(1535)}$ (GeV) & $ 1.530 $& $ 1.542 $& & \\
$\Gamma_{S_{11}(1535)}$ (MeV) & $ 142 $& $ 162 $& & \\[1ex]
 Mass of the third $S_{11}$ (GeV)&  & ~1.729~$\pm$~0.003& &  \\
 Width of the third $S_{11}$ (MeV)&  &183~$\pm$~10& &  \\[1ex]
$\chi^2_{d.o.f}$ &\multicolumn{1}{c}{$3.8$}  & \multicolumn{1}{c|}{$1.6$}& \multicolumn{1}{c}{}& \multicolumn{1}{c}{}    \\
\hline
\end{tabular}
\end{center}
\end{table}

In Figures 1 and 2,  we show comparison for the total and differential cross-sections,
respectively, for the two models I and II (Table~2).

Both models include all known resonances, given in Table~1, as well as those
with mass higher than 2 GeV. They also satisfy the configuration mixing
relations, Eqs.~(\ref{eq:MixS}) and~(\ref{eq:MixD}).

For the model I, the extracted mixing angles, Table~2, are in agreement with
Isgur-Karl~[9] predictions. The mass and the width of the $S_{11}(1535)$
come out in the PDG~[15] ranges. This model reproduces fairly well the total
cross-section data (Fig.~1) up to $W \approx$ 1.61 GeV. Between this latter energy and 
$\approx$ 1.68~GeV, the model overestimates the data, and above 1.68~GeV, the
predictions underestimate the experimental results, missing the total cross-section
increase. 

In summary, results of the model I show clearly that an approach containing a
correct treatment of the Born terms and including {\it all known resonances} in the 
{\it s-} and {\it u-}channels {\it does not} lead to an acceptable model, even within
broken $SU(6)\otimes O(3)$ symmetry scheme.
\begin{figure}[t]
   \begin{center}
   \includegraphics[height=22pc,width=40pc]{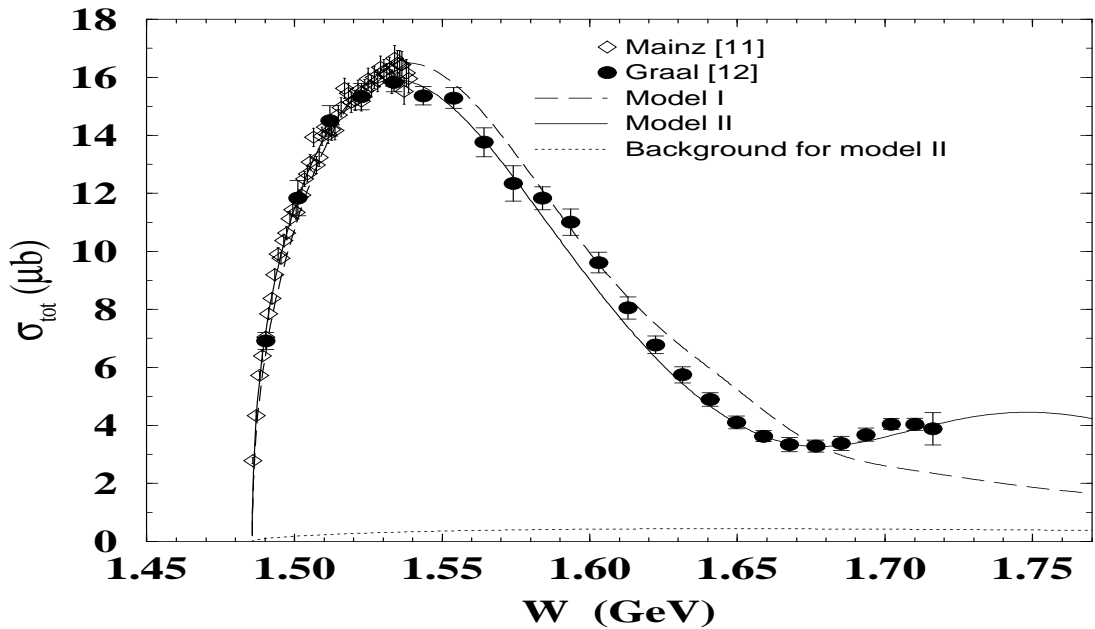}
   \end{center}
   \caption{Total cross section for the reaction 
$\gamma p \to \eta p$
as a function of total center-of-mass energy. 
The curves come from the models I (dashed), II (full). 
The dotted curve shows
the background terms contribution in the model II.
Data are from Refs.~[11]  (empty diamonds), and [12] (full circles).} 
\end{figure}

To go further, one possible scenario is to investigate manifestations of yet undiscovered
resonances, because of their weak or null coupling to the $\pi N$ channel.
A rather large number of such resonances has been predicted by several authors~[16-18].
To find out which ones could be considered as relevant candidates, we examined
the available data.

The excitation functions (Fig.~2), show clearly that this mismatch is due to the
forward angle peaking of the differential cross-section
for $W \ge$1.68~GeV ($E_{\gamma}^{lab} \ge$ 1.~GeV). Such a behaviour might
likely arise from missing strength in the $S$-waves. This latter conclusion is
endorsed by the role played by the $E_0^+$ in the multipole structure of the differential
cross-section and the single polarization observables.
If there is indeed an additional $S$-wave resonance in this
mass region, its dependence on incoming photon and outgoing meson momenta would be 
qualitatively similar to that of the $S_{11}(1535)$, even though the form factor
might be very different. Thus, for this new resonance, we use the same CGLN
amplitude expressions as for the $S_{11}(1535)$.
We have hence introduced~[1] a third $S_{11}$ resonance and refitted the same data base
as for the model I, leaving it's mass and width as free parameters. 
The results of this model, depicted in Figs.~1 and 2 (full curves), reproduce nicely the
data. This is also the case~[1] for the polarized beam and polarized
target asymmetries. For this latter observable, our predictions come out
in agreement with the data.
\begin{figure}[t]
   \begin{center}
   \includegraphics[height=30pc,width=40pc]{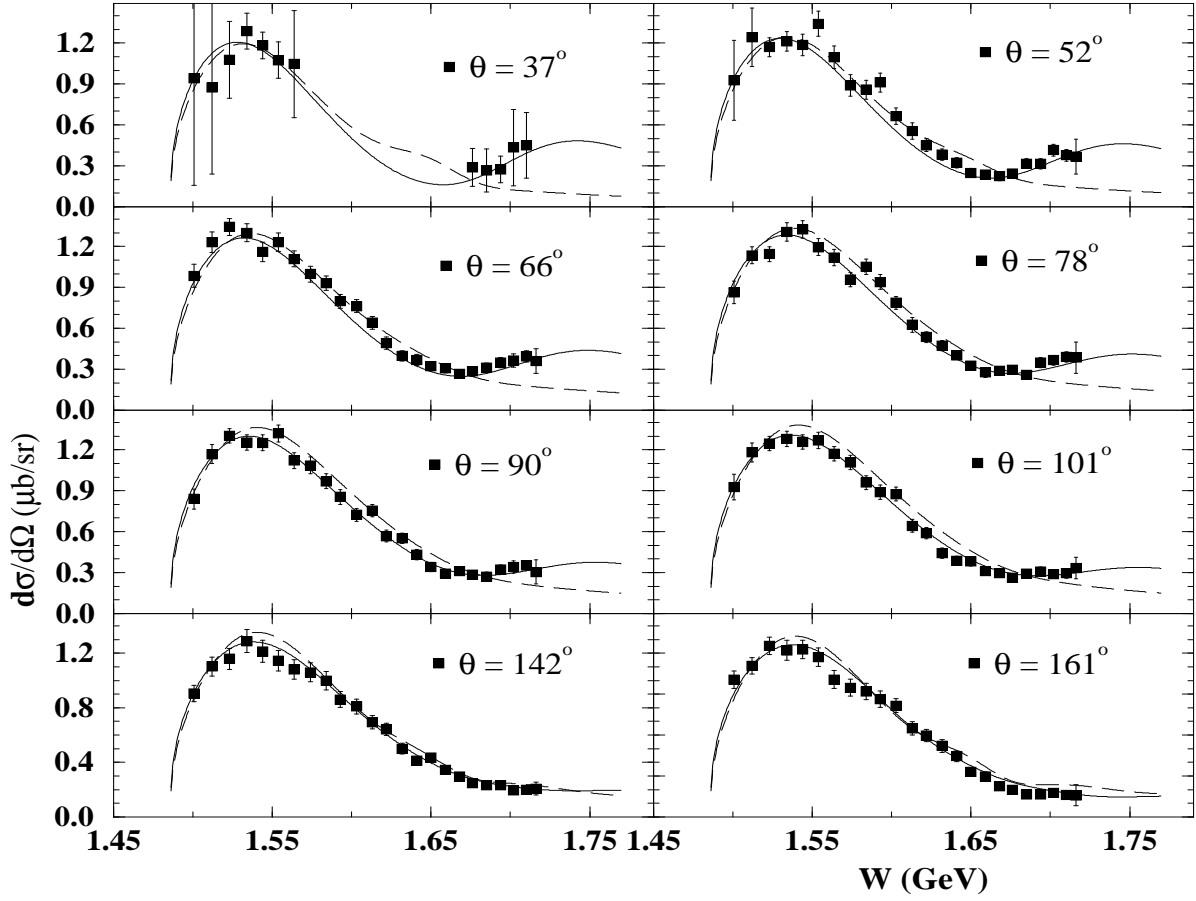}
   \end{center}
   \caption{Excitation function for the reaction 
$\gamma p \to \eta p$
as a function of total center-of-mass energy. 
The curves are as in Fig~1 and have been calculated at the angles given
in the figures.
Data are from Graal~[12] and correspond roughly to the given angle
$\pm$2$^\circ$.} 
\end{figure}

The extracted
values for different free parameters in the model II are given in Table~2, 
third column (Model II). The mixing angles are still compatible with the
quark model predictions~[9] and results coming from the 
large-$N_c$ effective field theory based 
approaches~[19-20]. The mass and the width of the $S_{11}(1535)$ are within the
ranges reported in the PDG~[15]. For the new $S_{11}$ resonance, we find
M=1.729 GeV and $\Gamma$=183 MeV. These values are amazingly close to those of
a predicted~[16] 
third $S_{11}$ resonance, with M=1.712 GeV and $\Gamma_{T}$=184 MeV.
Moreover, for the one star $S_{11}(2090)$ resonance~[15], the Zagreb group
coupled channel analysis~[21-22] produces the following values
M = 1.792 $\pm$ 0.023 GeV and $\Gamma_T$ = 360 $\pm$ 49 MeV.

Introducing this third resonance, hereafter referred to as $S_{11}(1730)$, modifies
the extracted values for the parameters of the two other $S_{11}$ resonances~[1].
The mass and width of the first $S_{11}$ resonance come out compatible with
their recent determination by the CLAS collaboration~[23],
as well as with those of the Zagreb group coupled channel analysis~[21-22].

In summary, a new $S$-wave nucleonic resonance is needed to interpret 
the recent $\eta$-photoproduction data between threshold and 
$E_{\gamma}^{lab} \approx$1.1 GeV. The crucial questions then are:
i) what is the nature of this resonance? ii) are there other relevant
reactions to be investigated?

With respect to the first question, the authors of Ref.~[16] suggest a
$K \Sigma$ molecular structure for the third resonance that they predict.
We need hence, to find out whether that resonance and the $S_{11}(1730)$
found here, are the same, or this latter resonance has a 3-quark structure.
One way is to go from the photoproduction to the electroproduction of 
$\eta$-meson. We are currently extending our electroproduction study~[4],
limited to the $S_{11}(1535)$ region data~[23], to higher W. Actually, 
the very recent higher energy data~[24] from JLab allow us to perform such 
investigations. The $Q^2$ dependence of the cross-section is expected to 
teach us about the nature of this resonance. It is worthwhile underlining
that those data, at the lowest measured $Q^2$, show also a minimum around
the same W as in Fig.~1. Moreover, if the $S_{11}(1730)$ has an exotic
$K \Sigma$ structure, strangeness production~[5] close to threshold
would deserve special attention from experimentalists.
Vector meson channels~[6,25] might also be of interest.

We hope that the investigation of pseudoscalar and vector mesons
electromagnetic production within the same chiral constituent quark
model, will offer an appropriate means in search for new resonance
and will allow us to deepen our understanding of the baryons spectroscopy.
\section{Acknowledgements}
I wish to thank the organizers for their kind and generous
invitation to this very stimulating and friendly meeting.
I am indebted to Zhenping Li for the 
results reported in this paper and coming from
enlightening collaboration.
I would also like to thank Qiang Zhao for fruitful collaboration.
\section{References}
\vspace{1pc}
\re
1.\ Saghai B., Li Z.\ 2001, 
nucl-th/0104084, {\it submitted to} Eur. Phys. J.
\re
2.\ Li Z., Saghai B.\ 1998,
Nucl. Phys. A {644}, 345
\re
3.\ Zhao Q., Didelez J.P., Guidal M., Saghai B.\ 1999,
Nucl. Phys. A 660, 323
\re
4.\ Zhao Q., Saghai B., Li Z.\ 2000,
{\it submitted to} Phys. Rev. C, nucl-th/0011069
\re
5.\  Li Z., Saghai B., Ye T.\ 2001, 
{\it in progress}
\re
6.\ Zhao Q., Saghai B., Al-Khalili J.S.\ 2001,
nucl-th/0102025, {\it to appear in } Phys. Lett. B,  
\re
7.\ Manohar A., Georgi H.\ 1984, 
	{Nucl. Phys.} B {234}, 189
\re
8.\ Li Z., Ye H., Lu M.\ 1997, 
	Phys. Rev. C {56}, 1099
\re
9.\ Isgur N., Karl G.\ 1977,
{Phys. Lett. B} {72}, 109
\re
10.\ Isgur N., Karl G., Koniuk R.\ 1978,
       {Phys. Rev. Lett.} {41}, 1269
%
%
\re
11.\ Krusche B.\ et al.\ 1995,
Phys. Rev. Lett. {74}, 3736
\re
12.\ Renard F.\ et al.\ 2000,
{\it submitted to} Phys. Rev. Lett., 
hep-ex/0011098.
\re
13.\ Ajaka J.\ et al.\ 1998,
Phys. Rev. Lett. {81}, 1797
\re
14.\ Bock A.\ et al.\ 1998,
Phys. Rev. Lett. {81} 534
\re
15.\ Particle Data Group\ 2000,
Eurp. Phys. Jour. {15}, 1
\re
16.\ Li Z., Workman R.\ 1996,
Phys. Rev. C {53}, R549
\re
17.\ Capstick S., Roberts W.\ 2000,
Prog. Part. Nucl. Phys., {45} (Suppl. 2), 5241
\re
18.\ Bijker R., Iachello F., Leviatan A.\ 2000,
Ann. Phys. {284}, 89
\re
19.\ Carlson C.E.\ et al.\ 1999,
Phys. Rev. D {59}, 114008; and references therein
\re
20.\ Pirjol D., Yan T.-M.\ 1998,
Phys. Rev. D {57}, 5434; and references therein
\re
21.\ Batini\'{c} M.\ et al.\ 1998,
Phys. Scripta {\bf 58}, 15
\re 
22.\ \v{S}varc A., Ceci S.\ 2000,
nucl-th/0009024
\re
23.\ Armstrong D.\ et al.\ 1999,
Phys. Rev. D {60}, 052004
\re
24.\ Thompson R.\ et al.\ 2001,
Phys. Rev. Lett. 86, 1702
\re
25.\ Zhao Q.\ 2001,
Phys. Rev. C {63}, 025203
%
\end{document}